\newcommand{\ie}{{\it i.e.}}
\newcommand{\eg}{{\it e.g.}}

\newcommand{\etal}{{\it et~al.}}

\newcommand{\rhs}{r.h.s.}
\newcommand{\BR}[1]{\linebreak[0]#1\linebreak[0]}
\newcommand{\onlinecite}[1]{\cite{#1}}

\documentstyle[twocolumn,prb,aps]{revtex}

\draft

\begin{document}

\title{
  Neighbor-junction state effect on the fluxon motion in a Josephson stack
}

\author{
  E.~Goldobin\cite{gold-email}
}

\address{
  Institute of Thin Film and Ion Technology,
  Research Center J\"ulich GmbH (FZJ) \\
  D-52425, J\"ulich, Germany
}

\author{
  A.~V.~Ustinov
}

\address{
  Physikalisches Institut III,
  Uni\-ver\-si\-t\"at Er\-lan\-gen-N\"urn\-berg,
  D-91054, Erlangen, Germany
}

\date{\today}

\wideabs{ 

\maketitle

\begin{abstract}

  We study experimentally and theoretically the influence of phase-whirling (resistive) state in one junction of a two-fold Josephson stack on the fluxon motion in the other junction. In experiment, we measure the fluxon velocity versus current in one junction as a function of the state (Meissner or resistive) of the neighboring junction. The analysis, made for the limit of high fluxon density, shows that the interaction with the resistive state results in an increase of the effective damping for the moving fluxon and, therefore, in reduction of its velocity. Numerical simulations confirm this result for various fluxon densities. The experimental data are in good agreement with the theoretical predictions. In addition, the fluxon step measured experimentally has a rather peculiar structure with back and forth bending regions which is understood as a manifestation of the photon absorption in the neighboring junction.

\end{abstract}


} 

\section{Introduction}

Stacked long Josephson junctions (LJJ's) have recently received much attention since they show a variety of new physical phenomena \cite{SBP,PUPS,CurLock:Cryogen92,Cherry1,CurrLock} in comparison with single LJJ's and have potential for applications as a narrow linewidth powerful oscillators for mm and sub-mm wavebands\cite{Ustinov:InPhaseModeStkSim}. The naturally layered high-$T_c$ superconductors (HTS) can be described as intrinsic stacks of Josephson junctions \cite{Intrinsic}. Therefore, study of fluxon dynamics in artificial stacks can help to understand the phenomena which take place in HTS.

The inductive coupling model describing the dynamics of Josephson phases in $N$ inductively coupled LJJ's was derived by Sakai \etal \cite{SBP}. Experimental investigation of stacked junctions became possible after the progress achieved in (Nb-Al-AlO$_{\rm x}$)$_N$-Nb technology \cite{TechSLJJ10} which, at the present stage, allows to fabricate stacks with up to about 30 Josephson tunnel junctions having parameter spread between them of less than 10\% \cite{Thyssen:PhD-Thesis}. Initially, the interest was concentrated on investigation of the simplest symmetric fluxon states since they are  promising for oscillator applications. Later on, it was found that it is very interesting to understand the {\em asymmetric} states because they show rather nontrivial nonlinear dynamics\cite{Cherry1,CurrLock}. Such asymmetric states are also of practical importance, because multilayered oscillators most probably will operate in a regime when only the majority but not all of the junctions oscillate coherently while the other junctions are in resistive or not synchronized flux-flow state\cite{Ustinov:InPhaseModeStkSim}.

In a recent work \cite{CurrLock} it has been shown that the dynamic state of one junction in a 2-fold stack affects the {\em static} properties of the other junction. As a next step, it is interesting and important to understand how different dynamic states in one LJJ affect the {\em dynamics} of fluxons in the other LJJ. In particular, the goal of this work is to study the dynamics of a fluxon in one LJJ when the neighboring LJJ is in the resistive state and compare it with the case when the neighboring LJJ is in Meissner state. We call the resistive state a ``phase-whirling'' state because the Josephson phase difference rotates very fast and nearly uniformly, in rough approximation. Such a dynamic state often occurs in experiments and, therefore, it is important to understand and describe it adequately. In fact, in the early experiments with stacks it was somewhat naively supposed that the voltage of Flux-Flow Step (FFS) in one LJJ does not depend on the state (Meissner or resistive) of the other LJJ. In fact this is true only for the {\em asymptotic} voltage of FFS. Here we show that in the presence of the ``phase-whirling'' solution in one of the junctions the actual flux-flow voltage across the other LJJ gets lower.

In the section II we present the experimental data which clearly show that in a 2-fold stack the switching of one junction from the Meissner state to the phase-whirling state decreases the velocity of a fluxon moving in the other junction and, therefore, the dc voltage across it. The analytical approach which explains the observed decrease of fluxon velocity in the limit of high fluxon density is developed in section III. The results of numerical simulations confirm analytical results and are shown in section IV. The results of the work are summarized in section V.

\section{Experiment}

In order to investigate the influence of the phase-whirling state in the neighboring junction on the fluxon dynamics, we have chosen the most clean ring-shape (annular) LJJ stack geometry. Due to magnetic flux quantization in a superconducting ring, the number of fluxons initially trapped in each annular junction of the stack is conserved. The fluxon dynamics can be studied here under periodic boundary conditions which exclude possible complicated interference of the fluxon with the junction edges.

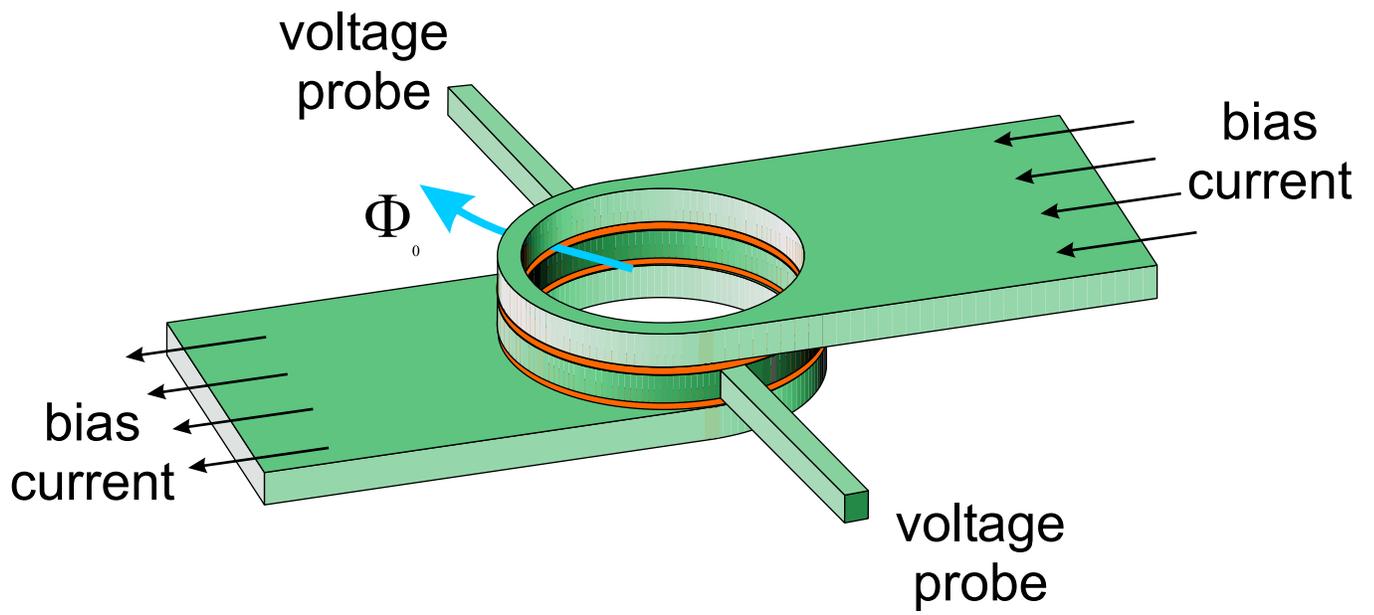
\begin{figure}
  \caption{
    Two coupled (stacked) long Josephson junctions of annular geometry with one fluxon trapped in the top junction. For symmetry, we used two voltage probes attached to the middle superconducting electrode. Dimensions are not to scale.
  }
  \label{Fig:Geometry}
\end{figure}

Experiments have been performed with 3 different (Nb-Al-AlO$_{x}$)$_2$-Nb annular LJJ stacks prepared in 2 different technological runs (2 samples in one run and the third sample in another run). The sample geometry is shown in Fig.~\ref{Fig:Geometry}. Two annular LJJ's are stacked one on top of the other, with bias leads attached to the top and bottom electrodes. The physical parameters of all samples, measured at $T=4.2\,{\rm K}$, are summarized in Tab.~\ref{Tab:Params}. The stacks were designed with extra contacts to the middle superconducting electrode \cite{Tech:StackWithMiddleElectrod} so that the voltages across each LJJ can be measured separately. The inner diameter of all stacks was $D=122.5\,{\rm{\mu{}m}}$ and the width $W=10\,{\rm{\mu{}m}}$. Due to technological difficulties of making a stack of identical LJJ's with contacts to the middle electrode, the two stacked junctions had rather substantial difference in quasiparticle (subgap) resistance $R_{\rm QP}$. The normalized circumference of the ring was $\pi D/\lambda_J=L/\lambda_J\approx 15$, where $\lambda_J$ is the Josephson penetration depth, which was approximately equal in both junctions. Measurements were performed in the temperature range $4.2$--$5.8\,{\rm{K}}$.

In stacked annular LJJ's, clean trapping of a single fluxon in a desired junction is rather difficult due to the asymmetry of the required state $[1|0]$. In the particular case of the 3 samples mentioned above, the asymmetry in the junction's resistance allowed to trap the fluxon in the desired $[1|0]$ state without many efforts just by applying a small bias current through one of the junctions during cooling the sample below the critical temperature $T_c$. After every trapping attempt, the resulting state was checked. The $I$--$V$ characteristic (IVC) of both LJJ's were traced simultaneously in such a way that the current was applied through the whole structure (through two junctions connected in series) and the voltages were measured individually across each LJJ. The wanted state $[1|0]$ with a fluxon in one junction and no fluxon in the other junction was identified by a simultaneous observation of a small critical current $I_c$ and fluxon step with the smallest asymptotic voltage $\sim20\,{\rm \mu V}$ in LJJ$^A$, and a large critical current in LJJ$^B$. Both the current amplitude of the fluxon step $I_{\max}^A(H)$ and the critical current $I_c^B(H)$ are expected to have their maxima at zero applied magnetic field $H=0$. To check that we have clean fluxon trapping, \ie{} that the fluxon is trapped in a LJJ and not accompanied by the parasitic Abrikosov vortices in the superconducting films surrounding LJJ, we checked the dependences $I_c^{A,B}(H)$ and $I_{\max}(H)$ after each trapping attempt and repeated it until these dependences were symmetric.

The main experimental result of the paper is shown in Fig.~\ref{Fig:ExpIVC}. It is IVC's of both LJJ's of the sample \#2 traced at $T\approx5\,{\rm K}$ using rather complex current sweep sequence. Note, that the voltage scales of two IVC's in Fig.~\ref{Fig:ExpIVC} are different and shown on the bottom axis for $V^A$ and on the top axis for $V^B$. The sweep starts at the bias point A where $I=0$ and $V=0$ and a fluxon is trapped in LJJ$^A$ (state $[1|0]$). When the current is increased up to about $I=0.69\,{\rm mA}$ (point B in Fig.~\ref{Fig:ExpIVC}), LJJ$^A$ switches to the fluxon step, while LJJ$^B$ still remains in the Meissner state. Ideally, the LJJ$^A$ should switch to the fluxon step at zero bias current since any non zero current applied should drive the fluxon around the stack. In our case the fluxon is pinned, most probably near one of the contacts to the middle electrode, and only current $I=0.69\,{\rm mA}$ can tear it away from the pinning center. With the further increase of the bias current the LJJ$^A$ follows the fluxon step which corresponds to the fluxon rotating in the ring, and the voltage across LJJ$^A$ is proportional to the fluxon rotation frequency according to the Josephson relation.

In an ideal single annular LJJ, the fluxon step has a relativistic nature and its slope approaches infinity when the fluxon moves with velocity $u$ close to the Swihart velocity ${\bar c}_{0}$. In the stack with different $j_c$ or with inhomogeneities, the fluxon's velocity can exceed the Swihart velocity. This results in the emission of the electromagnetic waves travelling behind moving fluxon (Cherenkov radiation) and in a finite slope of the step at any velocity\cite{Cherry1,Cherry2}. If the length of the emitted radiation tail is comparable with the circumference of the LJJ, the resonant structures (small steps) can appear on the top of the fluxon step. Such steps are visible on the top of the fluxon step in Fig.~\ref{Fig:ExpIVC} and are outlined by the circle. At $I\approx2.67\,{\rm mA}$ (point C in Fig.~\ref{Fig:ExpIVC}) both junctions simultaneously switch to the resistive state (gap voltage) with rapidly whirling Josephson phase. Such a simultaneous switching is called current locking. We studied it in detail for stacks of linear geometry in Ref.~\onlinecite{CurrLock}. Analyzing the dependence of critical current $I_c^B(H)$ and maximum current of the fluxon step $I_{\max}^A(H)$ on magnetic field, we conclude that the current locking was driven (initiated) by LJJ$^A$ (if LJJ$^A$ is kept in the resistive state, $I_c^B$ is substantially higher).

When both LJJ's are in the resistive state, we inverse the direction of the sweep, \ie{} start reducing the bias current. At $I=1.08\,{\rm mA}$ (point D in Fig.~\ref{Fig:ExpIVC}), LJJ$^A$ switches from the resistive state to the fluxon step while LJJ$^B$ still stays in the phase-whirling state. We denote such state of the stack as $[1|R]$. The voltage $V^A$ in the $[1|R]$ is by about $16\%$ smaller than $V^A$ in the $[1|0]$ state at the same bias. In fact, this is one of key observations in our study.

At this point there are two possibilities: first, continue to decrease the bias current down to zero or, second, increase the current and trace up the single fluxon step for $[1|R]$ state.

If we continue decreasing the bias current, at $I=0.916\,{\rm mA}$ (point E in Fig.~\ref{Fig:ExpIVC}) LJJ$^B$ switches from the resistive state (McCumber branch) to the Meissner state and the overall state of the stack becomes $[1|0]$. This causes the voltage $V^A$ to increase and become equal to voltage of the fluxon step which we traced in the beginning of the bias current sweep. The fact that switching of LJJ$^B$ caused a voltage jump across LJJ$^A$ is marked in Fig.~\ref{Fig:ExpIVC} by dotted arrow. Thus, we demonstrated experimentally that the change in the state of LJJ$^B$ affects the velocity of fluxon moving in LJJ$^A$. Further decrease of the bias current results in the fluxon pinning at $I=0.470\,{\rm mA}$ (point F in Fig.~\ref{Fig:ExpIVC}) and in the zero voltage across both LJJ's.

The second possibility is, being in the bias point D, to increase bias current and trace the fluxon step of the $[1|R]$ state up. This step has rather peculiar shape as shown in Fig.~\ref{Fig:ExpIVC}. In addition to the common trend to have smaller voltage than the fluxon step in the $[1|0]$, the step in $[1|R]$ state bends back and forth which possibly implies some interesting physics behind it. We also noticed that as we trace both IVC's in $[R|R]$ state from bias point C down to bias point D and  then in $[1|R]$ state from point D up to bias point G, the voltage across the LJJ$^B$ has a small hysteresis at voltages equal to the sum of the gap voltages of the superconducting electrodes constituiting the LJJ$^B$. This small hysteresis is shown magnified in the inset of Fig.~\ref{Fig:ExpIVC}. The voltage across LJJ$^B$ is somewhat smaller in the $[1|R]$ state than in the $[R|R]$ state. As soon as LJJ$^A$ switches to the resistive state (point G in Fig.~\ref{Fig:ExpIVC} and dotted arrow in the inset), this difference vanishes.

We propose the following explanation for the observed back bending. As we increase current starting from point D up to $I\approx1.5\,{\rm mA}$, corresponding to nearly vertical slope of the fluxon step, the voltage $V^B$ increases from $55\,{\rm \mu{}V}$ up to $1.9\,{\rm mV}$, \ie{} approaches the gap voltage. The fluxon motion in LJJ$^A$, due to the coupling between the junctions, causes oscillations of Josephson phase and, therefore, of electric and magnetic fields in LJJ$^B$. This leads to photon-assisted tunneling (PAT) effect \cite{Werthamer:PAT,Hasselberg:PAT} in LJJ$^B$. This effect was earlier observed by Giaever \cite{Giaever:DetJosEff:1965} also using a stack of two junctions. The characteristic frequency $\omega$ of photons absorbed in LJJ$^B$ is equal to the fluxon rotation frequency in LJJ$^A$. Due to PAT, one expects to observe a step at the gap sum voltage decreased by $\hbar\omega/e=2V^A$. In the low bias region shown in the inset of Fig.~\ref{Fig:ExpIVC} the gap sum step is not well defined which results in somewhat weaker gap suppression result. In fact, the maximum suppression of the voltage $V^B$ we have found is about $20\,{\rm \mu{}V}$ that corresponds to the top of the back bending region at $I\approx2.6\,{\rm mA}$. Since the Josephson voltage in LJJ$^A$ is also about $20\,{\rm \mu{}V}$, its effect on $V^B$ is only about 50\% of the expected PAT step voltage change. Thus, as we increase current, the gap voltage in LJJ$^B$ decreases due to photon-assisted tunneling. The resulting decrease of fluxon step voltage of LJJ$^A$ is associated with the appearance of an additional dissipation channel due to PAT. Since the PAT step on IVC is limited in voltage (by $2V^A$) and in current amplitude ($\propto$ to the amplitude of the first Josephson harmonic which, in the case of fluxon motion, saturates at some bias), the bending to the right caused by Cherenkov radiation appears to be stronger at $I>2.6\,{\rm mA}$ so that the fluxon step of LJJ$^A$ gains the positive slope again. The differential resistance at the top the fluxon step in $[1|R]$ state is rather high and no resonances are observed. This picture is typical for fluxon with Cherenkov radiation tail moving in a media with high dissipation.

The negative bias part of the IVC reproduces all the features described above for the positive half except for the small hysteresis between bias point B and F. Very similar IVC's were found for other 2 measured samples. The minor difference was in the particular values of the bias current in the points B, C, D, E, F and G which were also dependent on $T$. All samples showed that the voltage $V^A$ of the fluxon step in the $[1|0]$ state is somewhat higher than the voltage of the same step at the same bias in the $[1|R]$ state. The fluxon step in the $[1|R]$ state showed back and forth bending for all samples. The hysteresis between points B and F was intersecting with the hysteresis between points D and E for some samples and temperatures, so we had to use even more complex sweep sequence in order to trace out all possible dynamical states.

\section{Theory}

The main objective of this section is to analyze the origin of the decrease in the fluxon velocity in one LJJ due to the switching of the neighboring junction into the resistive state. Here we use the standard RSJ model which does not take into account the dependence of the dissipation on voltage at $V\sim{}V_g\approx2.4\,{\rm mV}$. Thus the gap related effects like PAT discussed above are neglected.

The fluxon dynamics in the system under investigation can be described in the framework of the inductive coupling model\cite{SBP} which for the case of two coupled junctions takes the form:
\begin{eqnarray}
  \frac{\phi_{xx}}{1-S^2}-\phi_{tt}-\sin\phi - \frac{S\psi_{xx}}{1-S^2}
  &=& \alpha\phi_t-\gamma
  ; \label{Eq:CsG_1}\\
  \frac{\psi_{xx}}{1-S^2}-\psi_{tt}-\sin\psi - \frac{S\phi_{xx}}{1-S^2}
  &=& \alpha\psi_t-\gamma
  , \label{Eq:CsG_2}
\end{eqnarray}
where $\phi$ and $\psi$ are the Josephson phases across the LJJ$^{A,B}$, respectively, $-1<S<0$ is a dimensionless coupling parameter, $\alpha\ll{}1$ is the damping coefficient describing the dissipation in the system due to quasiparticle tunneling, and $\gamma=j/j_c$ is the normalized the density of the bias current flowing through the stack. The coordinate $x$ and time $t$ are measured, respectively, in units of the Josephson length $\lambda_{J}$ and inverse plasma frequency $\omega _{p}^{-1}$ of single-layer LJJ. Most of relevant parameters of the junctions, such as effective magnetic thicknesses and specific capacitances, for the sake of simplicity are assumed to be equal in both LJJ's.

To understand the origin of an additional friction force we start from unperturbed (without \rhs) Eqs.~(\ref{Eq:CsG_1}) and (\ref{Eq:CsG_2}) and use the force balance equations to derive the shape of the IVC. We do not directly solve Eqs.~(\ref{Eq:CsG_1}) and (\ref{Eq:CsG_2}), but, rather, use trial solutions for $\phi(x,t)$ and $\psi(x,t)$, which approximate the Josephson phase profiles in the states $[1|0]$ or $[1|R]$. The choice of the trial functions is suggested by the results of numerical simulations presented in the following section. To simplify the mathematics and concentrate attention on the physical sense, the dense fluxon chain approximation is used. For this case we adopt the following trial solutions:
\begin{eqnarray}
  \label{Eq:Trial_first}
  \phi(x,t) = H(x-ut) + A_{r,m} \sin\left[ H(x-ut) \right]
  ; \label{Eq:Trial_phi}\\
  \psi^m(x,t) = B_m\sin\left[ H(x-ut) \right]
  ; \label{Eq:Trial_psi_m}\\
  \psi^r(x,t) = \omega t - \frac{1}{\omega^2}\sin(\omega t)
              + B_r\sin\left[ H(x-ut) \right]
  , \label{Eq:Trial_psi_r}
  \label{Eq:Trial_last}
\end{eqnarray}
where $\psi^m$ and $\psi^r$ are phases in LJJ$^B$ in the Meissner state and resistive state, accordingly (these are the two cases which we are going to compare); $u$ is the velocity of the fluxon chain; $A_{r,m}$, $B_{r,m} \ll 1$ are the constants which we are going to determine for resistive and Meissner states, respectively; $H$ is the average normalized magnetic field in the LJJ$^A$:
\begin{equation}
  H = \frac{2 \pi N}{\ell}
  , \label{Eq:Def:H}
\end{equation}
where $N$ is the number of fluxons trapped in annular LJJ$^A$ (for $[1|0]$ and $[1|R]$ states $N=1$), and $\ell=L/\lambda_J$ is the normalized length of the junctions. Dense fluxon chain approximation implies that $H \gg 1$.

Substituting Eqs.~(\ref{Eq:Trial_first})--(\ref{Eq:Trial_last}) into Eqs.~(\ref{Eq:CsG_1}) and (\ref{Eq:CsG_2}), and using the following approximations (which are justified in our case)
\begin{eqnarray}
  \sin\phi  \approx \sin\left[ H(x-ut) \right]
  ; \label{Eq:approx_1}\\
  \sin\psi^m \approx B\sin\left[ H(x-ut) \right]
  ; \label{Eq:approx_sin_psi_m}\\
  \sin\psi^r \approx \sin(\omega t)
  , \label{Eq:approx_sin_psi_r}
\end{eqnarray}
we arrive to the equations from which we can determine $A_{r,m}$ and $B_{r,m}$. The final result for the $[N|R]$ (resistive) state is
\begin{eqnarray}
  A_r &=& - \frac{D}{H^2Q}
  ; \label{Eq:A_r}\\
  B_r &=& \frac{-S}{H^2Q}
  , \label{Eq:B_r}
\end{eqnarray}
where we introduced notations
\begin{eqnarray}
  D &=& 1-u^2(1-S^2)
  ; \label{Eq:D}\\
  Q &=& \left( 1-u^2 \right)^2 - u^4S^2
  . \label{Eq:Q}
\end{eqnarray}
Note, that both $D>0$ and $Q>0$ for $u<{\bar c}_{-}$.

The result for the $[N|0]$ (Meissner) state is
\begin{eqnarray}
  A_m &=& -\frac{D H^2 + 1-S^2}{H^2\left( Q H^2 + D \right)}
  ; \label{Eq:A_m}\\
  B_m &=& \frac{-S}{Q H^2 + D}
  . \label{Eq:B_m}
\end{eqnarray}

To calculate IVC, $\gamma(u)$, we write the force balance equation
\begin{equation}
  2 \pi N \gamma = F_{\alpha}^A + F_{\alpha}^B
  . \label{Eq:Newton}
\end{equation}
Here $F_{\alpha}^{A,B}$ are the friction forces which develop in LJJ$^{A,B}$. The expression for the friction force is well known from the perturbation theory \cite{McLoughlinScott}
\begin{equation}
  F_{\alpha} = \alpha \int_0^{\ell} \phi_x \phi_t \: dx
  , \label{Eq:F_alpha}
\end{equation}
where we will use $\ell=2 \pi N/H$ following from (\ref{Eq:Def:H}). Since we are interested in the average friction force to get an IVC, we have to perform friction force averaging in time:
\begin{equation}
  {\bar F}_{\alpha} = \frac{1}{T}\int_0^T F_{\alpha}(t) \: dt
  . \label{Eq:F_alpha_aver}
\end{equation}
Since there are two characteristic frequencies in the system, fluxon (Josephson) frequency and the phase-whirling frequency of the resistive state, we have to choose the averaging interval $T$ in Eq.~(\ref{Eq:F_alpha_aver}) so that it will contain an integer number of periods of each frequency \ie{} $T=2 \pi k/\omega=2 \pi m/Hu$, and $k$, $m$ being the integer constants. After averaging, we get the following expressions for friction forces
\begin{eqnarray}
  {\bar F}_{\alpha}^A &=& \pi N \alpha H u \left( A_{r,m}^2+2 \right)
  ; \label{Eq:FA}\\
  {\bar F}_{\alpha}^B &=& \pi N \alpha H u B_{r,m}^2
  . \label{Eq:FB}
\end{eqnarray}
All information about the actual state is contained in $A_{r,m}$ and $B_{r,m}$ calculated above for the Meissner and resistive state of the LJJ$^B$.

Finally, we insert (\ref{Eq:FA}) and (\ref{Eq:FB}) into the force balance equation (\ref{Eq:Newton}) and get IVC's
\begin{equation}
  \gamma_{r,m}(u)
  =\frac{\alpha Hu}{2}\left( A_{r,m}^2 + B_{r,m}^2 + 2 \right)
  . \label{Eq:IVC}
\end{equation}
Now we can prove that IVC for the $[N|R]$ state is shifted to the region of lower velocities in comparison with the IVC for $[N|0]$ state, \ie{} that
\begin{equation}
  \delta(u)=\gamma_r(u)-\gamma_m(u)>0
  \mbox{ for all } |u|<{\bar c}_{-}
  . \label{Eq:delta}
\end{equation}
Substituting $A_{r,m}$ and $B_{r,m}$ from Eqs.~(\ref{Eq:A_r}), (\ref{Eq:B_r}),  (\ref{Eq:A_m}), (\ref{Eq:B_m}) into the expression (\ref{Eq:IVC}) and using the obtained expressions for $\gamma_{r,m}(u)$ in (\ref{Eq:delta}) we get
\begin{equation}
  \delta(u)=\frac{\alpha u\left( X_1 H^2 + X_2 \right)}
  {2 H^3 Q^2 \left( H^2 Q + D \right)^2}
  , \label{Eq:delta(X)}
\end{equation}
where $X_1$ and $X_2$ are defined as
\begin{eqnarray}
  X_1 &=& 2 Q D \left[ S^2 -Q\left( 1-S^2 \right) + D^2\right]
  ; \label{Eq:X_1}\\
  X_2 &=& D^2\left( D^2 + S^2 \right) - \left( 1-S^2 \right)^2 Q^2
  , \label{Eq:X_2}
\end{eqnarray}

Obviously, Eq.~(\ref{Eq:delta(X)}) is positive when both $X_1$ and $X_2$ are positive. To prove the latter, we express $D^2$ as a function of $Q$ using Eqs.~(\ref{Eq:D}) and (\ref{Eq:Q})
\begin{equation}
  D^2 = Q\left( 1-S^2 \right) + S^2
  . \label{Eq:D(Q)}
\end{equation}
Substituting (\ref{Eq:D(Q)}) into (\ref{Eq:X_1}) and (\ref{Eq:X_2}) we get
\begin{eqnarray}
  X_1 &=& 4 Q D S^2 > 0
  ; \label{Eq:X_1a}\\
  X_2 &=& S^2\left[ 3Q\left( 1-S^2 \right) + 2 S^2\right] > 0
  . \label{Eq:X_2a}
\end{eqnarray}
Thus (\ref{Eq:delta}) is proved and $\gamma_r(u)>\gamma_m(u)$ for any $u<{\bar c}_{-}$.

This result is in agreement with our experiment and simulation (see the following section). From physical point of view, the origin of the effect lays in the difference between Eqs.~(\ref{Eq:approx_sin_psi_m}) and (\ref{Eq:approx_sin_psi_r}) where the main term depends on the state of LJJ$^B$ resulting in different phase profile and different friction force for $[1|0]$ and $[1|R]$ states. The IVC's of fluxon steps $u^A(\gamma)$ for the states $[1|0]$ and $[1|R]$ calculated using Eq.~(\ref{Eq:IVC}) are shown in Fig.~\ref{Fig:PertTheoryIVC}. According to the calculations presented above the difference $\delta(u)$  diverges as $u$ approaches ${\bar c}_{-}$. Actually, for this case of a single fluxon in our relatively long junction, the dense fluxon chain approximation is not fully valid so one needs to perform more exact analysis or numerical simulations. The region of validity of our approximation is $A_r<1$ and $B_r<1$. Using Eqs.~(\ref{Eq:A_r}) and (\ref{Eq:B_r}) for the same parameters as in Fig.~\ref{Fig:PertTheoryIVC} we get that our approximation is valid up to $u\approx0.809$ while ${\bar c}_{-}\approx0.816$.

\section{Simulation}

To check the limitations of the analysis presented above we performed direct numerical simulations. Our simulations show that the effect observed in experiment and explained in the framework of high fluxon density approximation exists for any fluxon densities and any velocity even very close to ${\bar c}_{-}$. Another advantage of the simulation over the analytical approach is that the simulation fully reproduce the dynamics of the inductive coupling model and, therefore, all the effects possible in its framework.

The numerical procedure works as follows. For a given set of LJJ's parameters we simulate the IVC of the system, \ie{} calculate $\bar{V}^A(\gamma)$ and $\bar{V}^B(\gamma)$ while increasing $\gamma$ from zero up to $1$. To calculate the voltages $\bar{V}^A(\gamma)$ and $\bar{V}^B(\gamma)$ for each value of $\gamma$, we simulate the dynamics of the phases $\phi^{A,B}(x,t)$ by solving the Eqs.~(\ref{Eq:CsG_1}) and (\ref{Eq:CsG_2}) with the periodic boundary conditions:
\begin{eqnarray}
  \phi  ^{A,B}(0,\tilde{t}) & = &\phi  ^{A,B}(\ell,\tilde{t}) + 2 \pi N^{A,B}\\
  \phi_{\tilde{x}}^{A,B}(0,\tilde{t}) &
  = &\phi_{\tilde{x}}^{A,B}(\ell,\tilde{t})
  \quad , \label{Eq:AnnBC}
\end{eqnarray}
numerically using an explicit method [expressing $\phi^{A,B}(t+\Delta t)$ as a function of $\phi^{A,B}(t)$ and $\phi^{A,B}(t-\Delta t)$], and treating $\phi_{xx}$ with a five point, $\phi_{tt}$ and $\phi_{t}$ with a three point symmetric finite difference scheme. Numerical stability was checked by doubling the spatial and temporal discretization steps $\Delta x$ and $\Delta t$ and checking its influence on the fluxon profiles and on the IVC. The discretization values used for simulation were $\Delta x = 0.01$, $\Delta t=0.0025$. After simulation of the phase dynamics for $T_0=20$ time units we calculate the average dc voltages $\bar{V}^{A,B}$ during this time interval as
\begin{equation}
  \bar{V}^{A,B}
  = \frac{1}{T}\int_0^T \phi^{A,B}_t(t) \:dt
  = \frac{\phi^{A,B}(T)-\phi^{A,B}(0)}{T}
  \quad . \label{Eq:V}
\end{equation}
For faster convergence, we use the fact that $\bar{V}^{A,B}$ does not
depend on $x$ and, therefore, we also take advantage of the spacial averaging of the phases $\phi^{A,B}$ in (\ref{Eq:V}).

When the values of $\bar{V}^{A,B}$ are found from (\ref{Eq:V}), the dynamics of the phases $\phi^{A,B}(x,t)$ is simulated further during $1.2\:T_0$ time units, the dc voltages $\bar{V}^{A,B}$ are calculated for this new time interval and are compared with the previously calculated values. We repeat such iterations further increasing the time interval by a factor 1.2 until the difference in dc voltages $|\bar{V}(1.2^{n+1}\:T)-\bar{V}(1.2^n\:T)|$ obtained in two subsequent iterations becomes less than a given accuracy $\delta V=10^{-3}$. The particular value of the factor $1.2$ was found to be quite optimal to provide fast convergence as well as more effective averaging of low harmonics on subsequent steps. Very small value of this factor, \eg{} $1.01$ can result in very slow convergence in the case when $\phi(t)$ contains harmonics with the period comparable or larger than $T$. Large values of the factor, \eg{} 2 or higher, will consume a lot of CPU time already during the second or third iteration even when the convergence is good. After the voltage averaging for current $\gamma$ is complete, $\gamma$ is changed by a small amount $\delta\gamma$ to calculate the voltages in the next point of the IVC. As initial conditions here we use a distribution of phases (and their derivatives) achieved in the previous point of the IVC.

An example of calculated IVC is shown in Fig.~\ref{Fig:SimIVC}. To trace both the Meissner and resistive states we use the following sweep sequence: $\gamma$ increases from 0 up to 1 with a step $\delta\gamma=0.01$, then decreases down to 0.5 with a step $\delta\gamma=0.01$, and further down with a step $\delta\gamma=0.002$ until the state $[N|R]$ is reached as shown in Fig.~\ref{Fig:SimIVC}. From this point, we either sweep further down to $\gamma=0$ or up to $\gamma=1$ until both junctions switch to the resistive state. The decrease of voltage in $[1|R]$ state in comparison with $[1|0]$ state is very clearly seen in Fig.~\ref{Fig:SimIVC}. We performed this kind of simulations for the wide range of the parameters \ie{} for $|S|=0.1$, $0.2 \ldots 0.5$, and $N=1$, $2 \ldots 5$ (in total 25 pairs of IVC's) and found similar IVC's in all cases. For relatively dense fluxon chain $N/L=1$ we also compared the IVC's obtained by means of numerical simulation with IVC's derived analytically and found a good agreement as shown in Fig.~\ref{Fig:PertTheoryIVC}. Small difference in the slope of analytical and numerical IVC's is related to final density of fluxons in simulation $H=2\pi$ while theoretical curve corresponds to $H \gg 1$.

The limitations of our model due to voltage independent loss term $\alpha$ prevented proper calculation of the upper part of the fluxon step in $[1|R]$ state. Here numerical curve shows a series of small voltage jumps (see Fig.~\ref{Fig:SimIVC}) in LJJ$^A$ that are not observed in experiment. These jumps are related to excitation of fluxon-antifluxon states in LJJ$^B$ at high voltages. Such states were not found in experiment, most probably due to the increased dumping in LJJ$^B$ at gap voltage.

\section{Conclusion}

We investigated experimentally and theoretically the motion of a fluxon in one of two magnetically coupled long Josephson junctions. Two different cases are studied: (1) when the neighboring junction (one which does not contain any fluxon) is in the Meissner state, and (2) when the neighboring junction is in a phase-whirling (resistive) state. We found that the phase-whirling state in LJJ$^B$ slows down the fluxon motion in LJJ$^A$ and results in a shift of the fluxon step to lower voltage. This effects is detected experimentally, reproduced in simulations based on the inductive coupling model, and derived analytically in the high fluxon density approximation. In addition, experiment shows quite peculiar back and forth bending of the fluxon step in $[1|R]$ state which we explain as a result of increased dumping due to photon assisted tunneling effect in LJJ$^B$. The results of our study are also relevant for characterization of stacked Josephson junctions with large number of layers.

\acknowledgments

We thank Norbert Thyssen for the sample fabrication. Partial support of this work by Deutsche Forschungsgemindschaft (DFG) is also acknowledged.



%
\begin{figure}
  \caption{
    Experimentally measured IVC of LJJ$^A$ containing a fluxon (open circles) and LJJ$^B$ (solid dots). The IVC's are plotted with different voltage scales: top axis shows the voltage across LJJ$^B$ which bottom axis --- across LJJ$^A$. This plot also shows a magnified view of the IVC corresponding to LJJ$^B$ at $V\approx2.2\,{\rm mV}$.
  }
  \label{Fig:ExpIVC}
\end{figure}
\begin{figure}
  \caption{
    IVC's of $[1|0]$ and $[1|R]$ states as predicted by analytical model Eqs.~(\ref{Eq:IVC}) (shown by lines) and obtained as a result of numerical simulation (symbols) for $S=-0.5$, $N=5$, $\ell=5$ ($H=2\pi$).
  }
  \label{Fig:PertTheoryIVC}
\end{figure}
\begin{figure}
  \caption{
    IVC's obtained by means of numerical simulation for $\ell=5$, $\alpha=0.05$, $N=1$, $S=-0.5$. The arrows show the directions of the sweep.
  }
  \label{Fig:SimIVC}
\end{figure}

\begin{table}[tbh]

  \begin{tabular}{rccc}
  sample & \#1 & \#2 & \#3\\
  \hline
  $V_g$ (mV) & 2.38/2.54 & 2.37/2.54 & 2.51/2.61\\
  $I_c$ (mA) & $\sim6.0$ & $\sim6.2$ & $\sim7$  \\
  $R_{\rm QP}$ ($\Omega$)& 0.6/5.1 & 0.4/4.0 & 1.1/5.4\\
  $R_N$ ($\Omega$)& 0.19/0.18 & 0.16/0.17 & 0.18/0.18\\
  $\Delta{}I_g$ (mA)& 10/12 & 9/13 & 10/10
  \end{tabular}
  \caption{
    Physical characteristics of stacks measures at $T=4.2\,{\rm K}$. Two numbers separated by slash, are related to top and bottom LJJ of the stack, respectively.
  }
  \label{Tab:Params}
\end{table}


\begin{references}

\bibitem[*]{gold-email}
  e-mail: e.goldobin@fz-juelich.de, homepage:
  http:\BR{//}www\BR{.}geocities\BR{.}com\BR{/}e\_goldobin

\bibitem{SBP}
  S.~Sakai, P.~Bodin, and N.~F.~Pedersen.
  J.~Appl. Phys. {\bf 73}, 2411 (1993).

\bibitem{PUPS}
  A.~Petraglia, A.~V.~Ustinov, N.~F.~Pedersen, S.~Sakai,
  J.~Appl.~Phys. {\bf 77}, 1171 (1995).

\bibitem{CurLock:Cryogen92}
  I.~P.~Nevirkovets, H.~Kohlstedt, C.~Heiden,
  ICEC Suppl., Cryogenics {\bf 32}, 583 (1992).

\bibitem{Cherry1}
  E.~Goldobin, A.~Wallraff, N.~Thyssen, and A.~V.~Ustinov,
  Phys. Rev. B {\bf 57}, 130 (1998).

\bibitem{CurrLock}
  E.~Goldobin and A.~V.~Ustinov,
  Phys. Rev. B {\bf 59}, 11532 (1999).

\bibitem{Ustinov:InPhaseModeStkSim}
  A.~V.~Ustinov, and S.~Sakai,
  Appl. Phys. Lett. {\bf 73}, 686 (1998).

\bibitem{Intrinsic}
  R.~Kleiner, F.~Steinmeyer, G.~Kunkel, and P.~M\"uller,
  Phys. Rev. Lett. {\bf 68}, 2394 (1992);\\
  %
  R.~Kleiner and P.~M\"uller,
  Phys. Rev.~B {\bf 49}, 1327 (1994).

\bibitem{TechSLJJ10}
  H.~Kohlstedt, G.~Hallmanns, I.~P.~Nevirkovets,
  D.~Guggi, and C.~Heiden,
  IEEE Trans. Appl. Supercond. {\bf 3}, 2197 (1993).

\bibitem{Thyssen:PhD-Thesis}
  N.~Thyssen, Ph.D. thesis,
  Universit\"{a}t Erlangen-N\"urnberg (1999).

\bibitem{Tech:StackWithMiddleElectrod}
  H.~Kohlstedt, A.~V.~Ustinov, F.~Peter,
  IEEE Trans. Appl. Supercond. {\bf 5}, 2939 (1995).

\bibitem{Cherry2}
  E.~Goldobin, A.~Wallraff, and A.~V.~Ustinov,
  cond-mat/9910234.

\bibitem{Werthamer:PAT}
  N.~R.~Werthamer,
  Phys. Rev. {\bf 147}, 255 (1966).

\bibitem{Hasselberg:PAT}
  L.-E.~Hasselberg, M.~T.~Levinsen, and M.~R.~Samuelsen,
  Phys. Rev. B {\bf 9}, 3757 (1974).

\bibitem{Giaever:DetJosEff:1965}
  I.~Giaever,
  Phys. Rev. Lett. {\bf 14}, 904 (1965).

\bibitem{McLoughlinScott}
  D.~W.~McLaughlin, A.~C.~Scott,
  Phys. Rev.~A {\bf 18}, 1652 (1978).

\end{references}
\end{document}